\documentclass[12pt,letterpaper]{article}
\usepackage[letterpaper]{geometry}
\usepackage{amssymb,amsmath}
\usepackage[round,authoryear]{natbib}
\usepackage[colorlinks=true, linkcolor=black, citecolor=blue, urlcolor=blue]{hyperref}
\usepackage{graphicx}
\usepackage{url}
\usepackage{float} % to use [H]
\usepackage[skip=0pt]{caption} % reduce space between figure and caption
\graphicspath{{Figures/}}

\newcommand{\bM}{\boldsymbol{M}}
\newcommand{\bC}{\boldsymbol{C}}

\usepackage{color}
\definecolor{blue}{RGB}{0,0,255}
\definecolor{red}{RGB}{255,0,0}
\definecolor{purple2}{RGB}{102,0,204}

\newcommand{\blind}{0}

\begin{document}

\def\spacingset#1{\renewcommand{\baselinestretch}%
{#1}\small\normalsize} \spacingset{1}

\if0\blind
{
  \title{The bixplot: A variation on the boxplot\\ 
         suited for bimodal data}
  \author{
  Camille M. Montalcini\thanks{Biological and Environmental 
  Sciences and Engineering Division, King Abdullah 
  University of Science and Technology, Thuwal, Saudi 
  Arabia} 
  \and Peter J. Rousseeuw\thanks{Section of Statistics and 
  Data Science, KU Leuven, Leuven, Belgium}
  }
  \date{May 4, 2026}
  \maketitle
} \fi

\if1\blind
{
  \phantom{abc}\\
  \bigskip
  \bigskip
  \begin{center}
    {\LARGE The bixplot: A variation on the boxplot}\\
    \vspace{2mm}
    {\LARGE suited for bimodal data\\}
  \bigskip \bigskip
  \large{May 4, 2026} 
  \vspace{25mm}
  \end{center}
} \fi

\begin{abstract}
Boxplots and related visualization methods are
widely used exploratory tools for taking a first 
look at collections of univariate variables.
In this note an extension is provided that is 
specifically designed to detect and display bimodality and
multimodality when the data warrant it. For this
purpose a univariate clustering method is constructed
that ensures contiguous clusters (meaning that no cluster 
has members inside another cluster), and such that each
cluster contains at least a given number of unique
members. The resulting bixplot display facilitates the 
identification and interpretation of potentially 
meaningful subgroups underlying the data. 
The bixplot also displays the individual data values, 
which can draw attention to isolated points.
Implementations of the bixplot are available in both 
Python and \textsf{R}, and their many options are 
illustrated on several real datasets. For instance,
an external variable can be visualized by color 
gradations inside the display.
\end{abstract}

\noindent {\it Keywords:}  
clustering; exploratory data analysis; 
graphical display; violin plot; visualization. 

\spacingset{1.45} 

%========================================================
\section{Background and motivation}
John Tukey's boxplot was a cornerstone of his monumental
project called \textit{Exploratory Data Analysis} 
developed in \citep{tukey1977}. His goal was to create a 
branch of statistics that put the data front and center, 
by starting the analysis with a set of tools that allow 
the data to reveal interesting properties and patterns
before carrying out any probability-based inference. 
Many of his tools were graphical. At that time 
computing power was limited, leading to a focus on simple 
methods. Some could even be applied by hand, 
such as the so-called stem-and-leaf plots.
But whereas some of these methods became obsolete 
when computing power grew exponentially, 
the boxplot is still ubiquitous and will likely remain
so in the foreseeable future.

\begin{figure}[!ht]
\centering
\includegraphics[width = 1.0\textwidth]
  {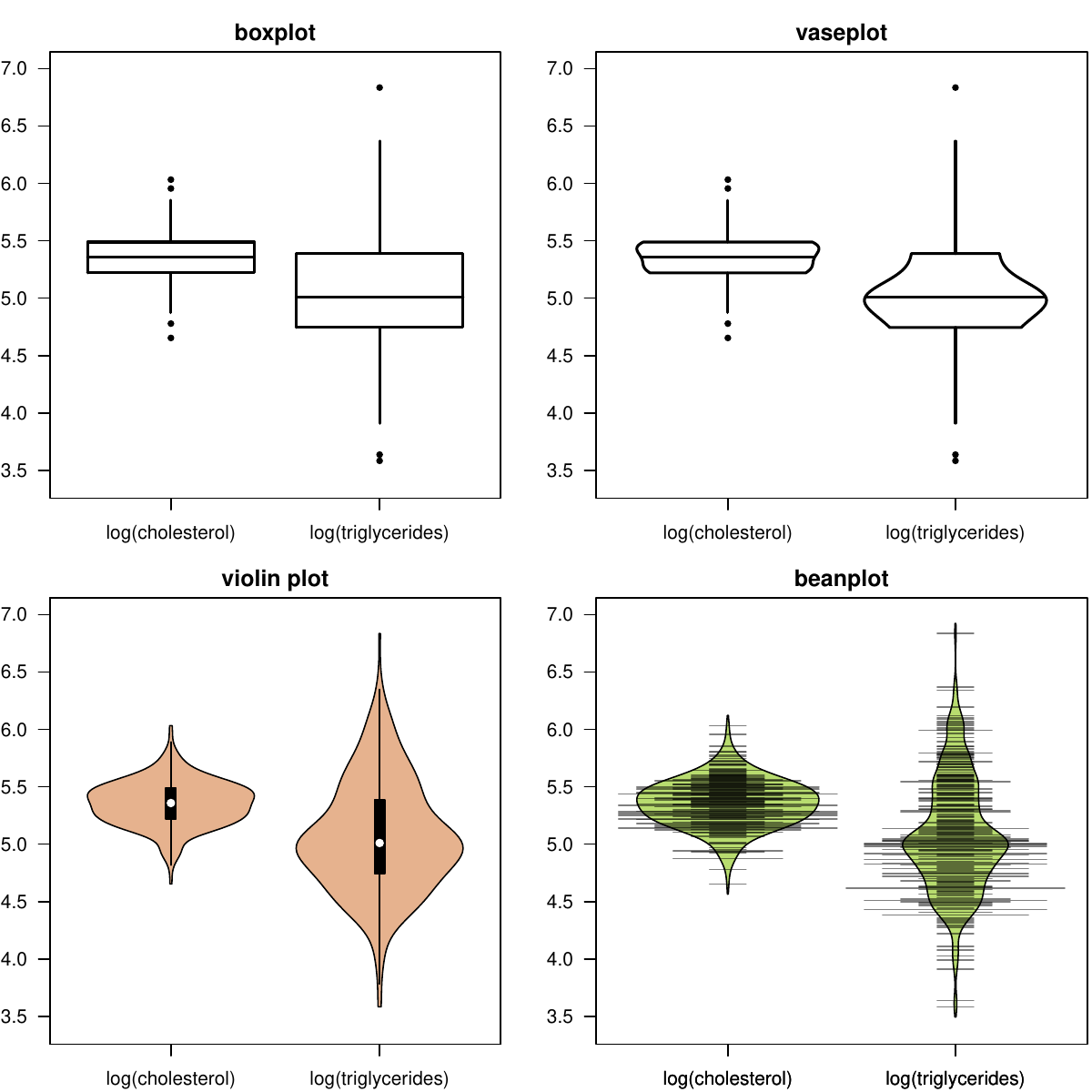}
\vspace{4mm}
\caption{Graphical displays of the bloodfat 
data from \citep{hand1994}.
Top left: classical boxplot of \cite{tukey1977},
top right: vaseplot of \cite{benjamini1988},
bottom left: violin plot of \cite{hintze1998},
bottom right: beanplot of \cite{kampstra2008}.}
\label{fig:historical}
\end{figure}

One might ask what gave the boxplot such staying power.
At the time, people seeing the boxplot for the first time
may have thought that it was merely an oversimplified
representation of a univariate distribution. Instead of
a histogram, there was only a box going from the first
quartile to the third quartile, inside of which a 
line indicated the position of second quartile, that is,
the median. For instance, look at the first boxplot
in the top left panel of Figure~\ref{fig:historical}.
The data are logarithms of the concentrations of 
cholesterol and triglycerides of 320 patients, analyzed 
by \cite{scott1978} and available in \citep{hand1994}.
The lines above and below the box are called the whiskers,
and range to the furthest datapoints that are considered 
inlying, according to a rule based on the quartiles.
Only the points outside these whiskers are actually
plotted, and flagged as outliers. Pointing
attention to outliers based on an objective rule was an 
instance of letting the data speak, and possibly tell 
the user something she might not have anticipated.

As a minimalist graphical display of a single univariate 
dataset the boxplot would not have gained many followers,
but its real strength is when several variables are
displayed simultaneously, as illustrated in the top
left panel of Figure~\ref{fig:historical}. In that 
situation the boxplots
allow easy comparison between the central location of 
these data sets (their median) and their scatter (the
height of the box, which is the interquartile range).
The position of the median inside the box says something
about symmetry, or lack of it. Finally, the whiskers and
plotted outliers depict the tails of the data distribution.
In this example we see that the central tendency of 
\texttt{log(triglycerides)} is lower than that of
\texttt{log(cholesterol)} and that its variability is
higher, that both samples look fairly symmetric, and 
that there are only a few points slightly outside the 
whiskers, meaning that there are no big outliers.
Nowadays, the results of numerical experiments are often
displayed by such parallel boxplots instead of tables.
Some extensions of the boxplot were proposed by 
\cite{McGill1978}, \cite{Bagplot1999} 
and \cite{hubert2008}. 

Over time, as more computing power became available,
various graphical refinements of the 1977 boxplot 
were proposed, roughly at decade-long intervals.
\cite{benjamini1988} introduced the 
\textit{vaseplot}, which replaced the sides of the
box by estimated density curves of the datapoints 
between the first and third quartiles, as in the 
top right panel of Figure~\ref{fig:historical}.
Next, \cite{hintze1998} proposed to plot the 
density over the entire data range, yielding the
\textit{violin plot} of the same data in the bottom 
left panel of Figure~\ref{fig:historical}. A narrow
version of the box and whiskers remains in the 
middle of each violin, without showing the outliers
explicitly. The final display in the bottom right 
panel of Figure~\ref{fig:historical} is the 
\textit{beanplot} of \cite{kampstra2008}. 
Unlike the previous displays it also shows the 
datapoints in the bulk of the data, as horizontal 
lines. Most of these have the same length, but when
several points coincide the short lines are stacked
by default. This draws attention to tied data values.
The stacking can be switched off when detecting ties
is not among the goals of the data analysis. 
The resulting set of horizontal lines, then all 
equally long, is called a \textit{rug plot} or a 
\textit{strip chart}.
The beanplot does not contain a box or whiskers,
but it allows to draw long horizontal lines at the 
average value or at the quartiles.
In \textsf{R} the boxplot is implemented as the
function \texttt{graphics::boxplot}, the violin plot
as \texttt{vioplot::vioplot} 
\citep{adler2025},
and the beanplot as \texttt{beanplot::beanplot}
\citep{kampstra2008}.
In Python, the library \texttt{seaborn}
\citep{Waskom2021} contains the functions 
\texttt{seaborn.boxplot()} and 
\texttt{seaborn.violinplot()}. All of these functions 
have an option to plot the data horizontally instead 
of vertically.

In spite of its many virtues, the boxplot also has some
deficiencies. It was noticed early on that the boxplot 
is unable to express that a dataset is bimodal. Indeed 
its design implicitly assumes that the dataset is 
unimodal, apart from possible outliers. 
Yet, bimodal univariate distributions are common 
across a variety of scientific domains, including 
psychology, medicine, ecology, and the study of animal 
behavior. In these fields, observed data often reflect 
a mixture of subpopulations that may arise from 
biological variability or heterogeneous experimental 
conditions. Recognizing and visualizing such distributional 
patterns can provide valuable insights into the underlying 
processes that generate them. For instance, reaction times 
in cognitive tasks can exhibit bimodality 
that reflect dual cognitive processes such as fast 
automatic versus slower controlled decision making 
\citep{freemanAssessingBimodalityDetect2013}. Behavioral 
and physiological traits can also display bimodal 
distributions, potentially reflecting phenotypic 
differences with a strong genetic basis.
While less common, multimodal distributions 
also occur. In ecology, for instance, 
individual body‑size distributions within bird 
communities have been shown to be  
multimodal, indicating clear ecological or functional 
groupings \citep{thibault2011}. In this note we 
will show bimodal and multimodal data involving 
fish, flowers, penguins, and cars.

Despite the ubiquity of non-unimodal distributions and 
the relevance of modes for data interpretation, there 
is currently no variation on the boxplot specifically 
designed to detect and display modes. The violin plot 
and the beanplot do share the important advantage that 
the plotted
density curve can have more than one local maximum,
which hints at non-unimodality, but no attempt is made
to fit a mixture of distributions to the data. Also,
the summary statistics in the violin plot are still
those of the boxplot inside of it: the median and
the first and third quartiles would not describe the 
data well if it came from such a mixture distribution, 
and the whisker endpoints based on them would be 
misleading. 

\begin{figure}[!ht]
\centering
\includegraphics[width = 1.0\textwidth]
  {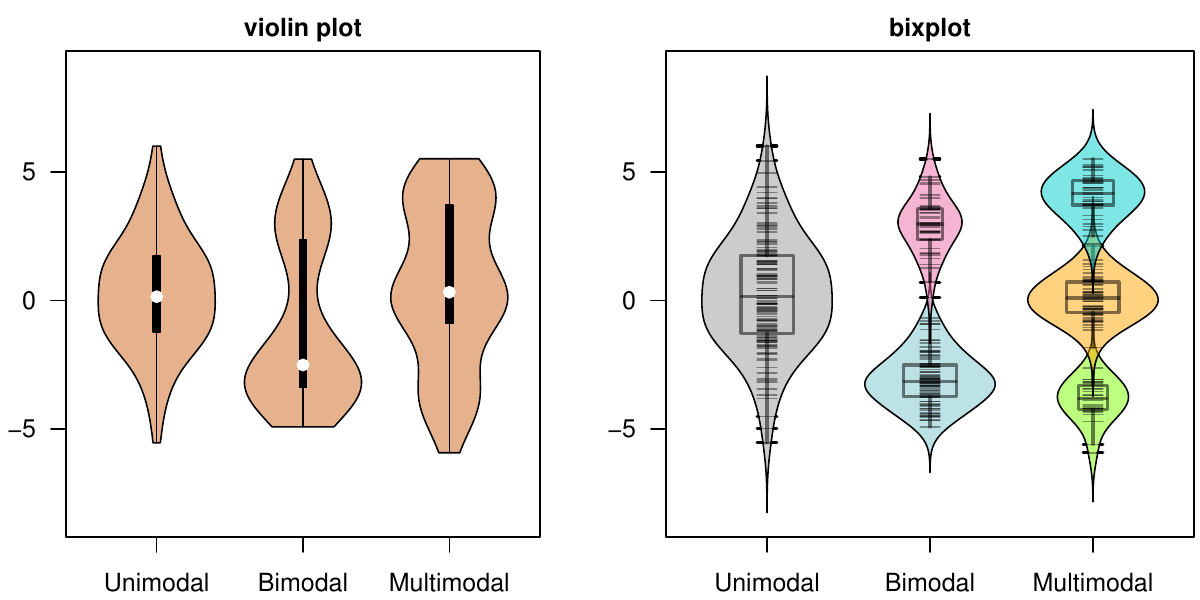}
\vspace{4mm}
\caption{Displaying generated unimodal, bimodal,
and multimodal variables by violin plots (left)
and bixplots (right).}
\label{fig:uni+bi+multimodal}
\end{figure}

To address this limitation we propose a different
variation on the boxplot that checks each variable for
unimodality. For variables deemed unimodal, the 
display combines a boxplot with elements of the violin 
plot and the beanplot. Variables not deemed unimodal 
are subjected to a clustering algorithm created for
this purpose, after which the clusters are displayed
with their boxes and estimated densities. 
Figure~\ref{fig:uni+bi+multimodal} visualizes
three generated datasets. The first is Gaussian,
the second is a mixture of two Gaussians with 
different sample sizes, and the third is a mixture
of three Gaussians. In the left panel we see that 
the fitted densities in the violin plots have the 
appropriate number of local maxima, but we do not
see the data inside the violins so it is hard to
separate the mixtures visually. 
The proposed plots are in the right hand panel.
The display of the unimodal data looks similar to
the violin plot but with a superimposed boxplot,
and the rug visualizes all the datapoints as in
a beanplot. The two components of the bimodal data
are now shown by separate bodies in different
colors, and in this example the densities 
overlap. The rug also offers visual support for 
the existence of two clusters. We can see that the 
bottom cluster contains more members, because here
the areas of the bodies are drawn proportional to 
the cardinality of the clusters (also other sizing
options are available). The three components of 
the multimodal data are separated in the same way.

Since the proposed visual display is suited 
for \textbf{bi}modal data and multiple 
m\textbf{ix}tures, we combined \textbf{bi}
and \textbf{ix} to the string \textbf{bix}
yielding the short name \textit{bixplot}. The name 
can be justified further by the fact that it combines 
the \textbf{b}o\textbf{x}plot, the density trace of 
the v\textbf{i}ol\textbf{i}n plot, and the rug of 
the \textbf{b}eanplot.
There are options for the color of the body
inside the density curve, the density curve itself, 
the border of the boxplot, the part of the rug lines
inside the body, and the part outside the body. The 
user can also set the line widths in all of these 
plots, as well as the degree of transparency in 
the body. The user can also choose whether to 
omit the body, density, boxplot, or rug plot.

The paper is structured as follows. 
Section~\ref{sec:examples} applies the bixplot display 
to several real datasets and illustrates some of its 
graphical options, making use of our implementations in 
both \textsf{R} and Python.
Section~\ref{sec:construction} sketches the main steps 
of its construction, and 
Section~\ref{sec:conclusions} concludes. 
The Supplementary Material contains more examples
and technical information about the algorithm.

\section{Real data examples} \label{sec:examples}

\begin{figure}[!ht] 
\centering
\includegraphics[width = 1.0\textwidth]
  {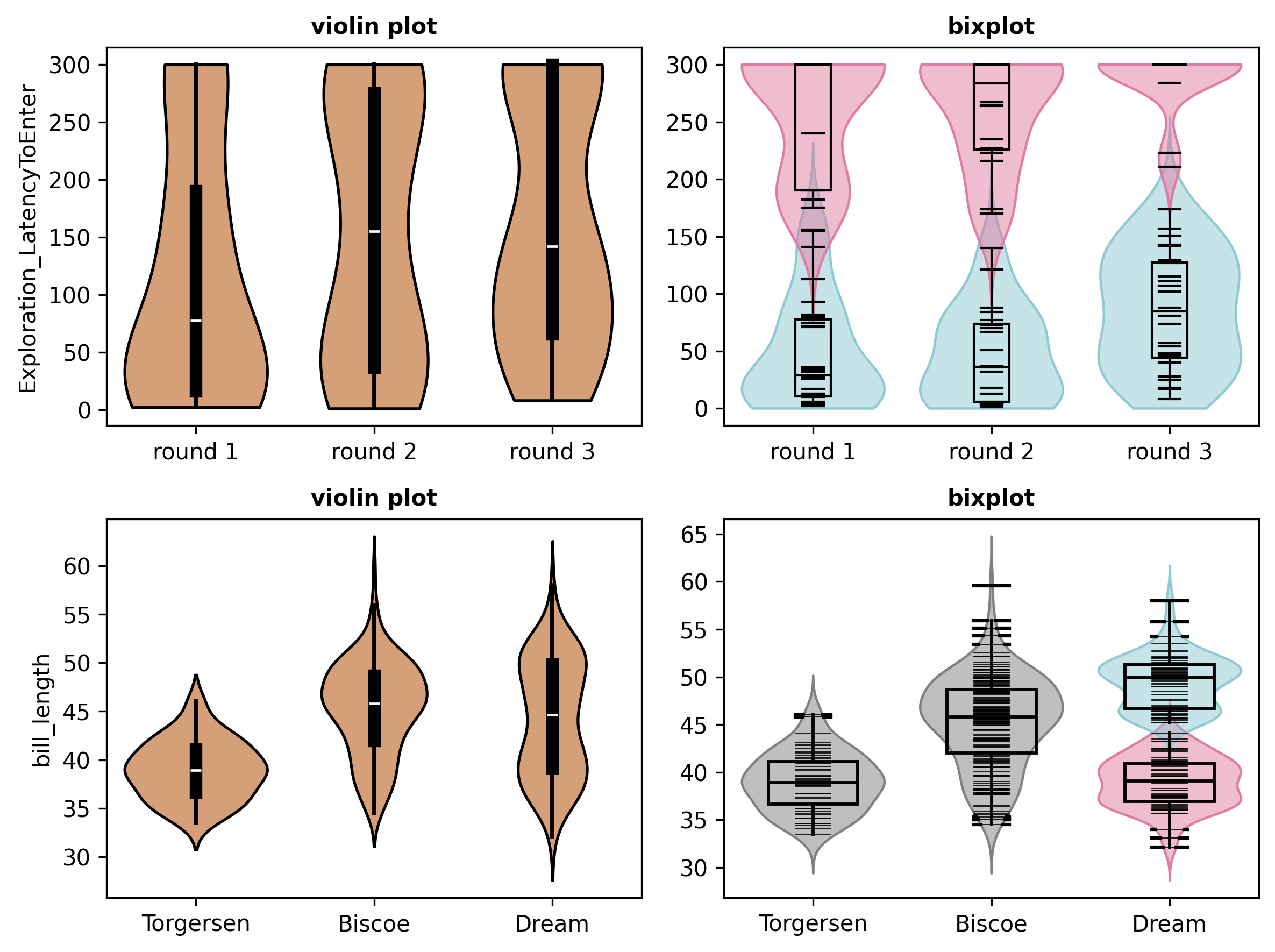}
\vspace{2mm}
\caption{Comparison of violin plots and bixplots for 
(top row) behavior latency time of fish, where the 
data are bounded by upper and lower limits; and 
(bottom row) bill length of penguins from three 
islands.}
\label{fig:latency+billength}
\end{figure}

The first dataset is about exploratory behavior in 
fish during three experimental rounds 
\citep{kerrPersonalityDoesNot2021}. The measurements 
are latency times in seconds. The top row of
Figure~\ref{fig:latency+billength} shows their 
violin plots on the left, which hint at bimodality
in each round. This is more pronounced in the bixplot 
display on the right. The output of the bixplot 
function provides summary statistics of each cluster, 
including the medians and quartiles corresponding to 
the boxplots inside. They indicate that the 
centers of the modes shifted upward in later rounds.
An option was used to restrict
the bixplot display to latencies between 0 seconds
and 300 seconds, because the dataset was bounded 
by these lower and upper limits.

The Palmer penguins dataset is available in 
\textsf{R} as \texttt{datasets::penguins} and was
analyzed in detail by \cite{Vanderplas2023}. 
It describes adult penguins covering three species 
found on three islands in the Palmer Archipelago, 
Antarctica, including their size (flipper length, 
body mass, bill dimensions) and sex. The bottom
row of Figure~\ref{fig:latency+billength} displays
the bill length measurements of penguins on the
three islands (Torgersen, Biscoe, and Dream).
The bixplot display on the right uncovers 
bimodality in animals from Dream Island but not 
from Biscoe Island, a distinction that the violin 
plot on the left did not make as clearly. The rug 
plot of Dream Island confirms the bimodality.

\begin{figure}[H] 
\centering
\includegraphics[width=1.0\textwidth]{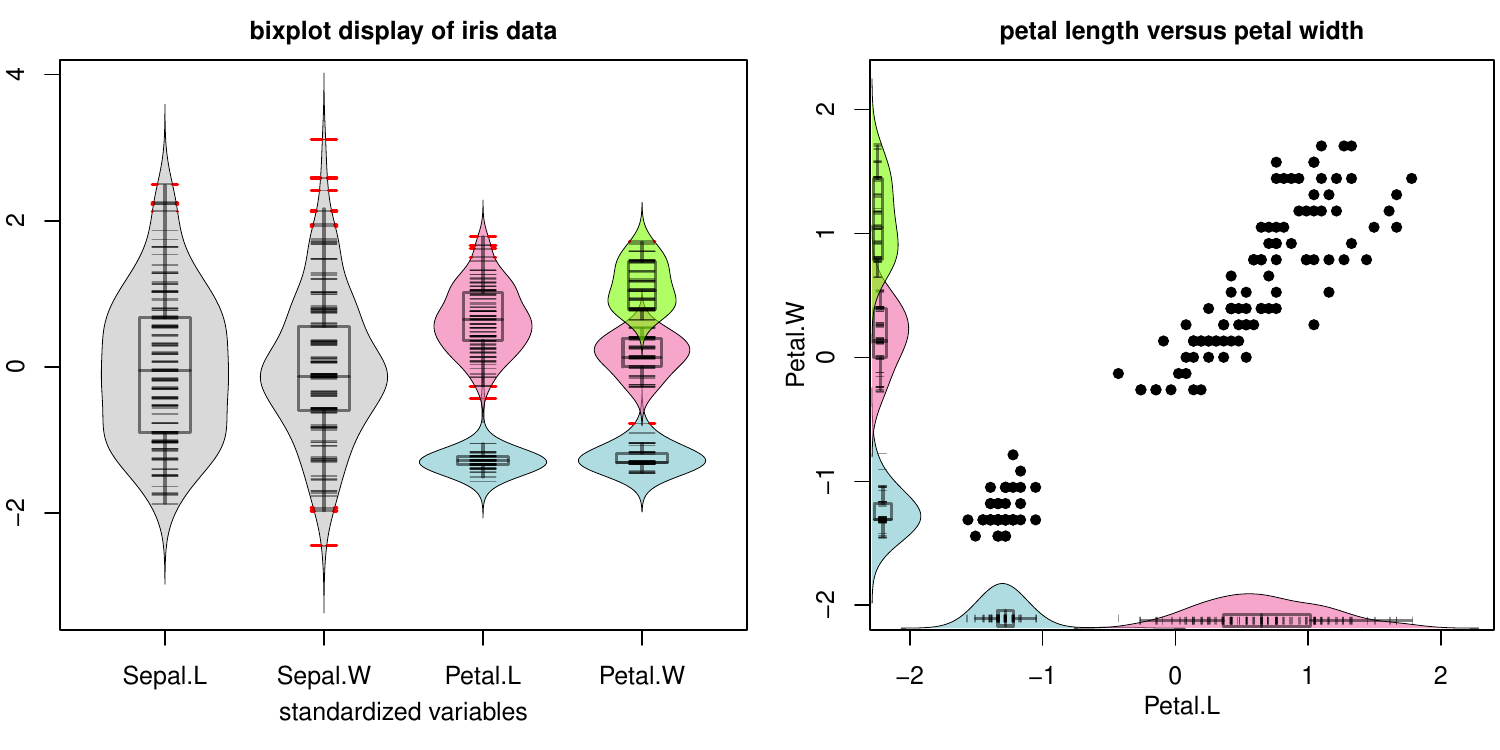}
\caption{(Left) bixplot of the standardized variables
of the iris data; (right) plot of the fourth versus
the third variable.}
\label{fig:iris}
\end{figure}

Another illustration of the bixplot as an
exploratory tool uses the well-known iris data.
It contains measurements of the variables sepal 
length and width, and petal length and width, for 50 
flowers from each of 3 species of 
iris. Let us pretend for a moment that we see these
data for the first time and that we have to analyze 
them without knowing the class labels (flower 
species). If we first standardize the four variables 
and run the bixplot function on them, we obtain the
left panel of Figure~\ref{fig:iris}. It tells us that 
the sepal length and width appear unimodal, whereas 
the petal length seems bimodal in the sense that it 
can be described by two clusters. The petal width 
might even be summarized by three clusters. As the 
latter two variables contain the most 
structure we plot them against each other, yielding 
the right panel of Figure~\ref{fig:iris}. 
Its marginal densities were also produced by the
bixplot function, using options to make the bixplots
one-sided and to add them to the existing scatterplot.
The points in the lower left part of the scatterplot
form a nicely separated 
cluster (which is actually the Setosa species). 
The remaining points look like a single cluster in
the petal length direction, while petal widths suggest 
they could have two modes. To resolve this ambiguity, 
the next step in the data exploration could be to 
apply a clustering algorithm to the multivariate
data formed by all 4 variables together. Depending on
the method used, this would separate these remaining 
points in two clusters that roughly correspond to the 
Versicolor and Virginica species. The Supplementary 
Material contains a similar display of the Top Gear 
car dataset of \cite{Alfons:robustHD}.

\begin{figure}[!ht] 
\centering
\includegraphics[width = 1.0\textwidth]{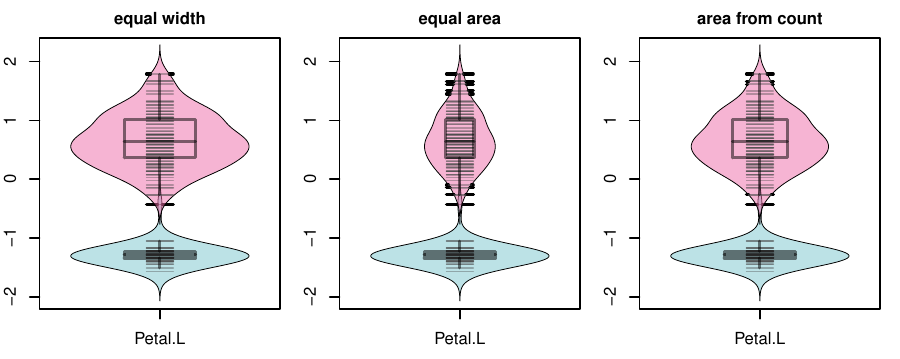}
\caption{Bixplots of the iris petal length using three 
sizing options for the bodies between density curves.
In the left panel each body has the same width.
In the middle panel the area of each body is the same. 
In the right panel the area of each body is 
proportional to the number of members in its cluster.}
\label{fig:sizing}
\end{figure}

The bixplot display offers many options to tailor
it to your needs. For instance, in the left panel 
of Figure~\ref{fig:iris} the option was chosen
to plot the parts of the rug lines outside of 
the bodies in bright red. This makes points in 
low-density regions more noticeable, such as the 
largest values of \texttt{Sepal.W}. Apart from 
outliers, also points that lie in between 
well-separated clusters stand out in this 
way, like the one in \texttt{Petal.L}. Moreover, 
we see that the unimodal 
variables \texttt{Sepal.L} and \texttt{Sepal.W}
have bodies of the same widths. That is the 
default, but it can be changed at will by the user.
Also the widths of each box and rug plot can be
chosen, as well as all the colors. The bodies of
the clusters of \texttt{Petal.L} and 
\texttt{Petal.W} have different colors, that can 
also be chosen. By default the areas of the bodies 
are proportional to the number of members in the 
clusters of the same variable, but there is 
also an option to make the areas of all clusters
of the same variable equal, or to equalize the 
widths of the bodies in the same variable.
Figure~\ref{fig:sizing} illustrates the three 
versions for the \texttt{Petal.L} variable.
The different sizing options affect the visual 
prominence of modes while preserving their 
summary statistics.
 
\begin{figure}[!ht]
\centering
\includegraphics[width = 1.0\textwidth]
  {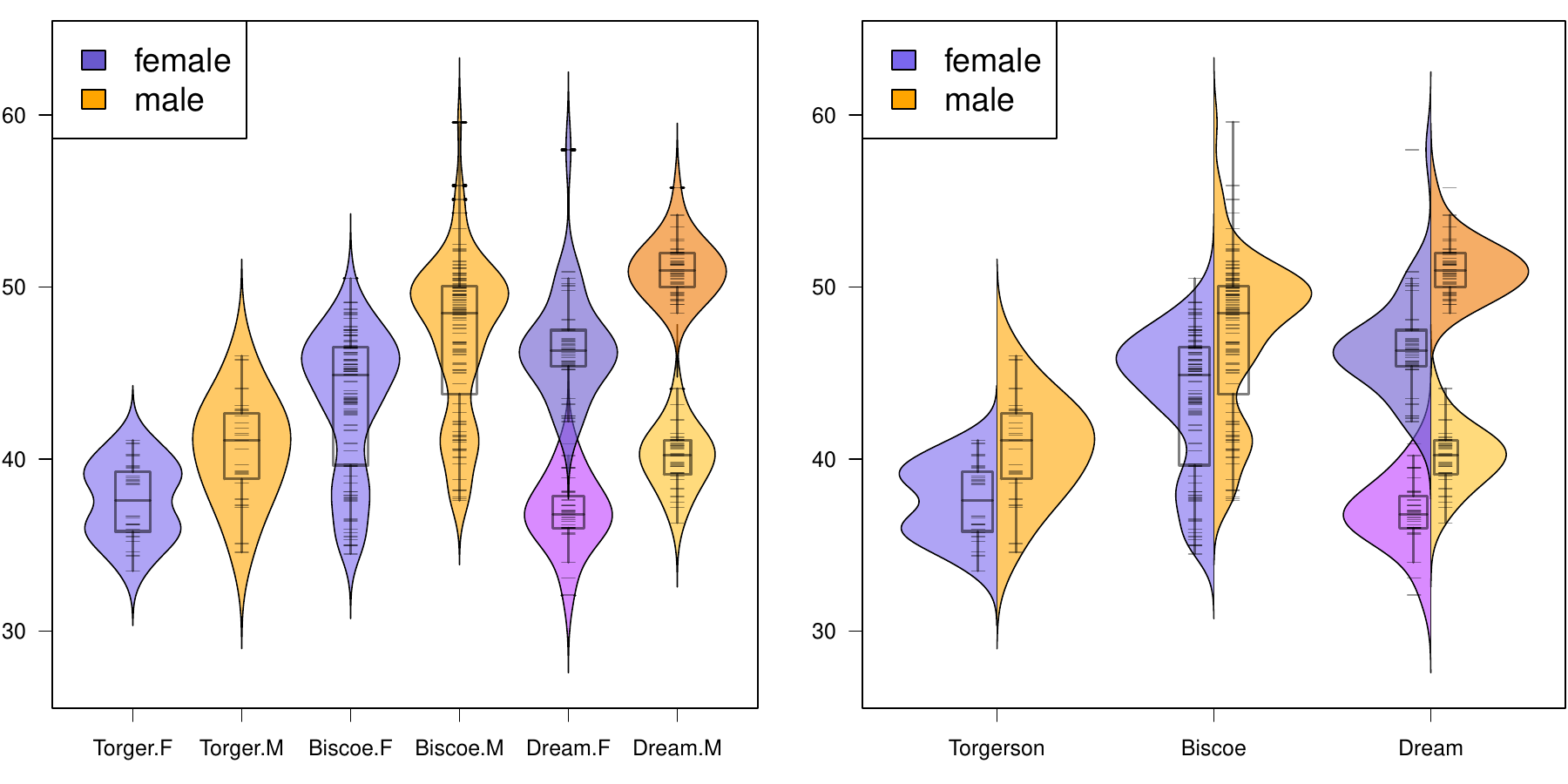} 
\caption{Penguins data: bixplots of bill length as
a function of island and sex. On the left we have
six bixplots on six vertical axes, and on the right 
the same results are represented by six `half' 
bixplots on three vertical axes, one for each island.}
\label{fig:billength_sex}
\end{figure}

We now return to the penguins data. The bottom row
of Figure~\ref{fig:latency+billength} displays the
bill length as a function of the island, but now
we would like to perform a more detailed analysis by
also taking the sex of the animals into account.
The left panel of Figure~\ref{fig:billength_sex}
has six bixplots, one for each combination of island
and sex. We see that the first four of them are
deemed to consist of a single cluster, whereas on
Dream Island both the female and male penguins show
two clusters. The display represents females in 
purple and males in orange. In this 
situation there is an option we can apply, 
that mirrors the bixplots of females and males on 
opposite sides of each island’s vertical line, 
as in the right hand panel of
Figure~\ref{fig:billength_sex}.
Each island thus shows two `half' bixplots, one for
each sex. This allows for direct within-island 
comparisons, and saves horizontal space.

\begin{figure}[!ht] 
\centering
\includegraphics[width = 1.0\textwidth]
   {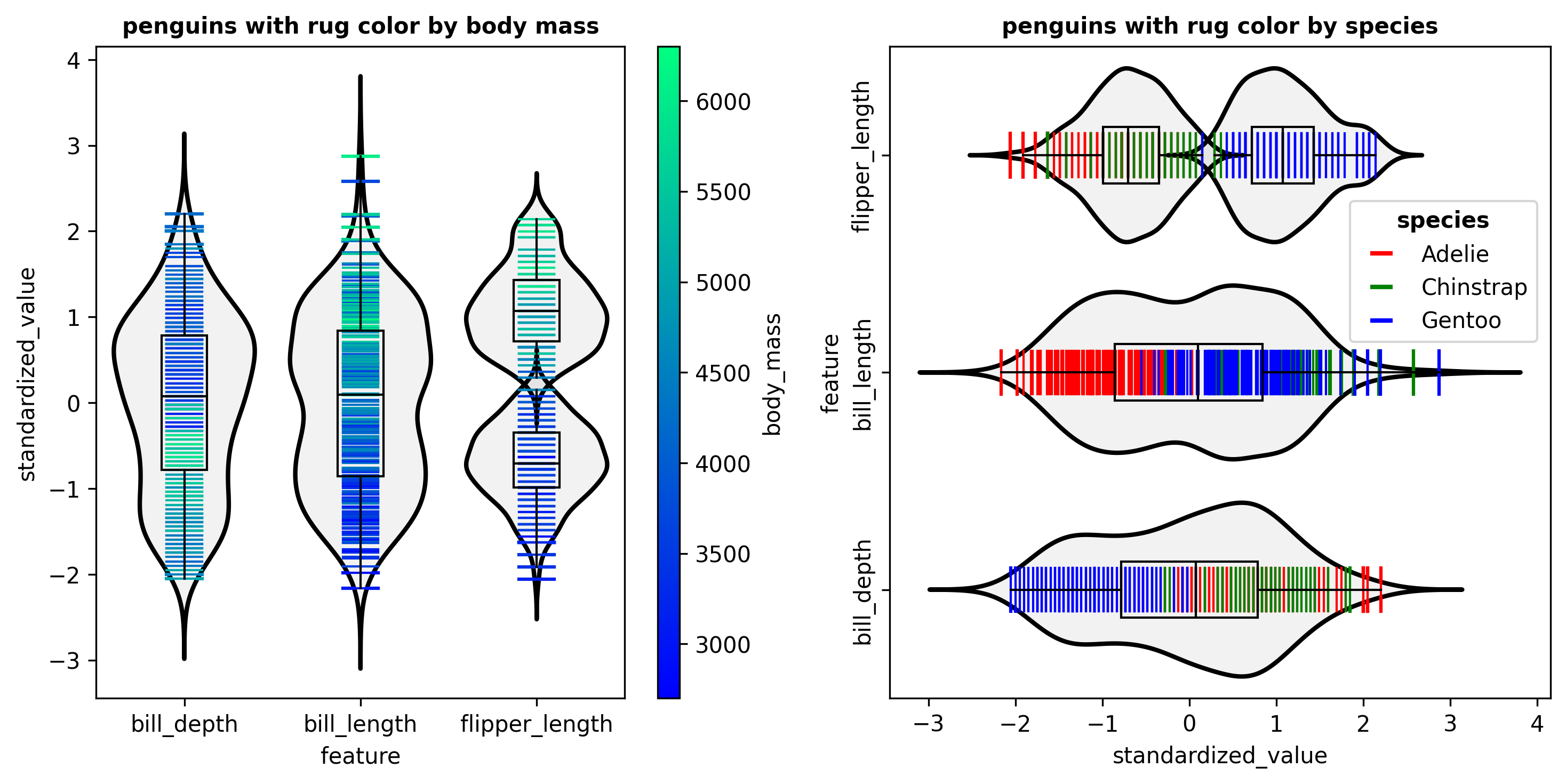}
\caption{Penguins data: coloring the rugs of bixplots 
visualizes relations with an additional variable.
Left panel: rug colors reflect the continuous variable
\texttt{body\_mass}. Right hand panel: rug colors 
reflect the categorical variable \texttt{species} in a
horizontal bixplot display.}
\label{fig:rugcolor}
\end{figure}

The left panel of Figure~\ref{fig:rugcolor} shows 
bixplots of the standardized \texttt{bill\_depth},
\texttt{bill\_length}, and \texttt{flipper\_length}
variables of the penguins data, this time not split
by island. What is new here is that the rugs inside
them do not have a single color. Instead the rugs are
colored according to another continuous variable, the
\texttt{body\_mass} of the animal, with its color
bar on the right. The display reveals that 
\texttt{body\_mass} has a strong positive 
association with the variable 
\texttt{flipper\_length}. Its relation with the other 
two variables is less clear and requires looking at
scatterplots.

The rug colors can also be mapped to a categorical 
variable. In the right hand panel of
Figure~\ref{fig:rugcolor}
the rug lines are colored according to the species
variable with levels Adelie, Chinstrap and Gentoo.
This panel also illustrates the option to plot the
bixplots horizontally.
We see at a glace that Gentoo penguins tend to have
a smaller \texttt{bill\_depth} and longer
\texttt{flipper\_length}.
In the Supplementary Material the rugs of the iris 
data are colored by their species too. There we 
also color the rugs of continuous variables in the 
Titanic data \citep{Friendly2019} by a categorical
variable.

The bixplots in this note and the Supplementary
Material can be drawn with both our Python and
\textsf{R} implementations. In fact, 
Figures~\ref{fig:latency+billength},
\ref{fig:rugcolor}, \ref{fig:kingfisher},
\ref{fig:topgear}, and \ref{fig:titanic} were 
made in Python, and the others in \textsf{R}. 
The plot commands can be very short due to their 
default options. We ensured that option names closely 
match those of \texttt{boxplot} in \textsf{R} and 
of \texttt{seaborn.violinplot} in Python, 
so existing calls to these functions will often run 
with \texttt{bixplot} as well.

%========================================================
\section{Construction of the bixplot} 
\label{sec:construction}

When constructing the bixplot of a variable, that is, a
univariate numeric dataset, the first thing we look at 
is the number $n$ of non-missing values. To reduce the 
risk that detected clusters arise from noise alone, we 
allow to search for at most \texttt{n/minN} clusters, 
where \texttt{minN} defaults to 15. So if $n$ is under
twice \texttt{minN} we do not attempt to cluster the data.
Also when the user specifies the maximal number of 
clusters \texttt{kmax} as 1, the data are treated as a
single cluster. In both cases the body of the bixplot
is determined by a kernel density estimate as in the
violin plot, and the boxplot and rug plot are superimposed.
There is an option to jitter the position of the rug 
lines, which is useful when the data contains many ties.

When $n \geqslant \texttt{2*minN}$ and 
$\texttt{kmax} \geqslant 2$ we test whether the 
hypothesis of unimodality should be rejected. For this 
purpose we use Hartigans' dip test \citep{hartigan1985}. 
The dip test detects a deviation from unimodality by 
computing the vertical distance between the empirical 
distribution function $\widehat{F}_n$ of the data and
the unimodal distribution function closest to 
it. Formally, the dip statistic 
$D(\widehat{F}_n)$ is given by
$$D(\widehat{F}_n) = \inf_{G \in \,\mathcal{U}} \,
  \sup_x |F(x) - G(x)|\;,$$
where $\mathcal{U}$ is the class of unimodal cumulative 
distribution functions. It turns out that the computation
of the dip statistic is not trivial, and neither is the
computation of its $p$-value. Fortunately we can 
use the fast implementations in the 
\textsf{R} package \texttt{diptest} \citep{diptest}
and in the Python package \texttt{diptest} 
\citep{diptestPython}.

If the p-value of $D(\widehat{F}_n)$ is below
a chosen level, the null hypothesis is rejected and the 
distribution is considered non-unimodal. By default we
use the 0.01 significance level.

When Hartigans' dip test rejects unimodality, we need 
to obtain an appropriate clustering of the data. In 
view of the interpretation of clusters as modes we 
need the clusters to be contiguous, that is, no cluster 
has members inside another cluster. In other words, the
clusters must be inside successive disjoint intervals. 
Not all methods of clustering or fitting mixture 
distributions satisfy this property. But clustering 
methods that assign each point to the cluster with the 
closest central point do satisfy it. We can compute 
central points by the robust $k$-medoids clustering 
method of \cite{kmedoids1987}. It partitions data
into $k$ clusters by selecting actual data points, 
called medoids, as cluster centers. It minimizes the 
total deviation $T$ defined as
the sum of distances between each point $x_i$ and the 
medoid $\mu_g$ of its assigned cluster $C_g$\,: 
$$T := \sum_{g=1}^{k} 
\sum_{x_i \in C_g} d(x_i, \mu_g)\,.$$
This is also called \textit{Partitioning Around
Medoids} (PAM). It is implemented in the 
\texttt{KMedoids} class of the Python library 
\texttt{scikit-learn-extra} \citep{scikit-learn-extra}, 
and as the \texttt{pam} function in the R package 
\texttt{cluster} \citep{clusterCRAN}. 

Moreover, in order to be able to interpret clusters as
modes we need to obtain clusters with sufficiently
many members. By itself the $k$-medoids method does not
guarantee this, and indeed a single isolated point may 
be put in a cluster by itself. This issue also occurs
with the $k$-means method, and was addressed by
\cite{bradley2000} who constructed a version of
$k$-means with the constraint that each cluster contain
at least a given number of members. Their algorithm
makes use of linear programming. We have extended this
idea to $k$-medoids and implemented it. We went a bit
further by constraining the clustering to having a 
minimal number \texttt{clusMinN} of \textit{unique} 
values in each cluster. This allows to draw a
meaningful density curve, and avoids the algorithm
trying to satisfy the constraint by assigning tied
cases to different clusters. By default 
\texttt{clusMinN} is set to 3, but the user can 
change this. Our
algorithm starts from the unconstrained $k$-medoids
method, and then each step reassigns points to 
clusters by a transport method, alternated
with recomputing the centers. These iteration steps 
cannot increase the objective function. The process 
continues until no further changes occur in the 
assignments or a given number of iterations is 
reached. A more thorough description of the
constrained $k$-medoids method is provided in
Section~\ref{suppmat:technical} of the Supplementary
Material.

The above algorithm is carried out for values of $k$
ranging from 2 to \texttt{kmax}. The maximal number
of clusters \texttt{kmax} is set by the user and
can be at most \texttt{floor(n/minN)}.
It remains to choose $k$. For each value of $k$ we 
compute the silhouette score \citep{Silhouettes1987} 
of the resulting clustering, given by
$$s(k) =\frac{1}{n} \sum_{i=1}^{n} 
\frac{b(i) - a(i)}{\max\{a(i), b(i)\}}$$
where $a(i)$ is the average distance of point 
$x_i$ to all other points in its cluster, and $b(i)$ 
is the lowest average distance of $x_i$ to any other 
cluster. We then select the number of clusters as the
$k$ yielding the highest $s(k)$. Note that the 
silhouette score is not defined for $k=1$. Therefore, 
this criterion cannot make a choice between $k=1$ and 
any $k \geqslant 2$. Fortunately, we already made 
that choice at the start, by Hartigans' dip test.

\begin{figure}[!ht]
\centering
\includegraphics[width = 1.0\textwidth]
  {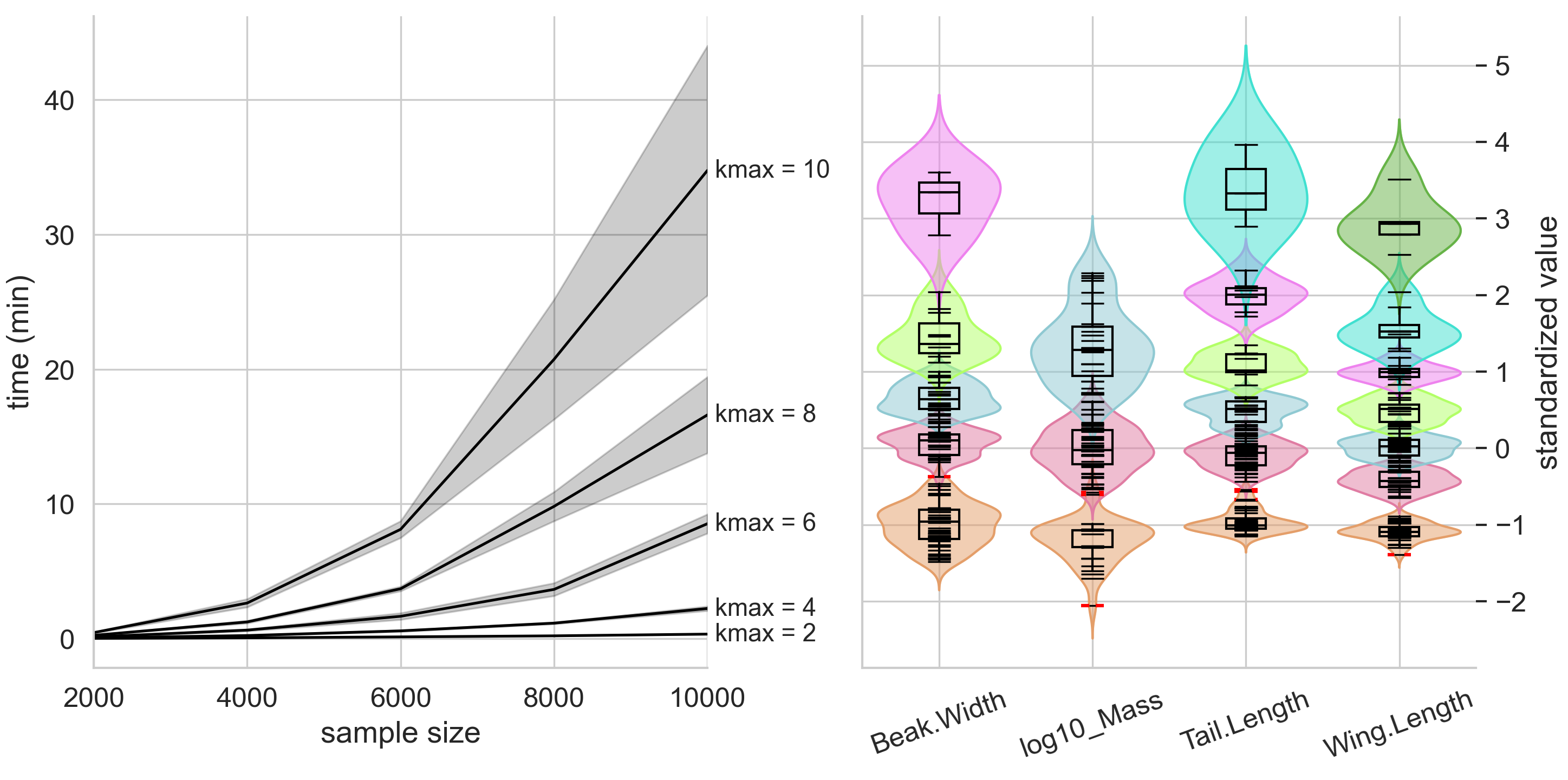} 
\caption{Computation times of the bixplot (left),
  and bixplot of the kingfisher data (right).}
\label{fig:kingfisher}
\end{figure}

The computational complexity of one 
iteration step of unconstrained $k$-medoids 
clustering is only \mbox{$O(n(n-k))$} if the 
fast swaps of \cite{FastPAM2021} are used, 
that are an option of \texttt{pam()} in 
the \textsf{R} package \texttt{cluster}. However,
an iteration step of the constrained $k$-medoids
clustering involves a call to the function 
\texttt{lp.transport} in the \textsf{R} package 
\texttt{lpSolve} \citep{lpSolve}, that uses the 
dual simplex method. Its average time complexity 
is $O((n + k)^3)$, so the constraint roughly
costs a factor $n$. In addition to this also
\texttt{kmax} plays a role, because $k$ ranges
from 2 to \texttt{kmax} so we need to carry out
\texttt{kmax} - 1 clusterings. The left panel 
of Figure~\ref{fig:kingfisher} shows the 
average computation times of bixplot running on 
artificial datasets with \texttt{kmax}
well-separated clusters, in function of $n$ and 
\texttt{kmax}. The gray area around a curve 
is $\pm 1$ standard deviation. The 
computation time indeed grows substantially,
and the resulting rugs were overplotted.
To avoid such long computation times,
when $n$ exceeds a given \texttt{bigN} the 
computations are carried out on a random subset 
of size \texttt{bigN}. The default \texttt{bigN} 
is 500, but the user can set it higher. The
code also puts a ceiling on \texttt{kmax} that 
is 7 by default, because we rarely encounter 
datasets with more than 7 clearly distinct
clusters, and we want to avoid overcrowded
bixplots anyway.

Multimodality with more than five clusters is 
rare in univariate real data. 
However, for illustrative purposes, one such example 
is provided by a dataset of 120 kingfisher birds
(\textit{alcedinidae}) from the AVONET database 
\citep{AVONET}, comprising the 
variables \texttt{Beak.Width}, \texttt{log10\_Mass}, 
\texttt{Tail.Length}, and \texttt{Wing.Length}. In 
the right hand panel of Figure~\ref{fig:kingfisher} 
we see that the clustering of \texttt{Wing.Length} 
identified 7 clusters, which are however difficult
to distinguish visually compared to other bixplots
in this paper. Here the multimodal structure could
be explained by different geographic origin.

%====================================================
\section{Conclusions} \label{sec:conclusions}
This note introduced the bixplot, an extension of
the boxplot that was specifically designed to detect 
and display clusters in univariate data and 
collections of univariate variables. For this 
purpose a dedicated constrained univariate 
clustering method has been developed. The bixplot 
display incorporates density curves as in the 
violin plot, as well as a rug plot that shows 
the individual data values as in the beanplot. 
In order to facilitate the use of the bixplot 
display, we have constructed extensive 
implementations in both Python and \textsf{R}, with 
practical defaults that allow for short function 
calls. On the other hand, a lot of options are 
available to tailor the display to the user's needs. 
Several real data illustrations have been provided 
in the text and the Supplementary Material.

We would like to stress that the bixplot display is
an exploratory tool, meaning that its purpose is to
assist the user's intuition about the data. It is 
not a confirmatory method like a significance test.
For instance, when the bixplot shows three clusters
in a variable this is an indication, and not a
confirmation that the population the data came
from has exactly three components. There exist
inferential methods for that purpose but they 
require probabilistic assumptions like 
Gaussianity, that we do not wish to restrict
ourselves to.\\

\noindent \textbf{Software availability.} The
Python implementation with examples is available 
in the \texttt{bixplot} package at 
\url{https://pypi.org/project/bixplot}\,.
The \textsf{R} implementation is the function
\texttt{bixplot()} in the \textsf{R} package
\texttt{classmap} \citep{classmap} at 
\url{https://CRAN.R-project.org/package=classmap}
 with the \mbox{vignette} 
\texttt{bixplot\_examples}\,.\\

\spacingset{1.1}
% \bibliographystyle{chicago}
% \bibliography{bib}

\clearpage

\pagenumbering{arabic}
% restarts page numbering from 1

\appendix

% \setcounter{figure}{7} 
% to continue numbering from the main text.
\renewcommand{\thefigure}{S\arabic{figure}}
\setcounter{figure}{0}

\begin{center}

\LARGE{\bf Supplementary Text}\\
\end{center}

\spacingset{1.45}

%====================================================
\section{More examples of the bixplot display}
\label{suppmat:examples}

\begin{figure}[H] 
\centering
\includegraphics[width = 1\textwidth]
  {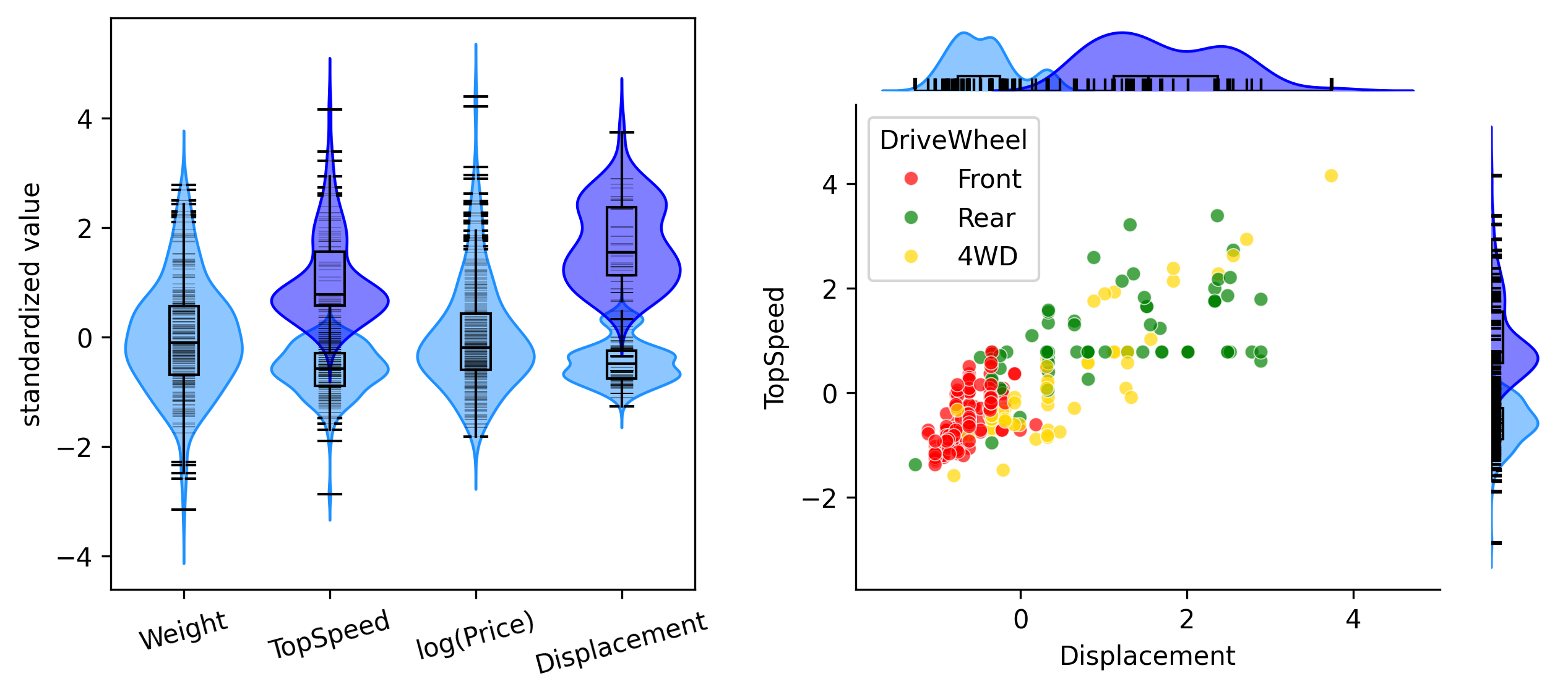}
\caption{(Left) bixplot of standardized variables 
of the Top Gear data; (right) plot of 
\texttt{TopSpeed} versus \texttt{Displacement}.}
\label{fig:topgear}
\end{figure}

The Top Gear data of Alfons (2016) contains
numerical and categorical variables on 297 cars.
Here we consider four numeric variables: the
\texttt{Weight} of the car, its \texttt{TopSpeed} 
and \texttt{Price}, and its engine 
\texttt{Displacement}, as well as the categorical 
variable 
\texttt{DriveWheel} with the levels \texttt{Front},
\texttt{Rear} and \texttt{4WD}. The variable
\texttt{Price} is very skewed so we log
transformed it. Afterward the four numerical
variables were standardized. The bixplot of these
variables revealed an outlier in \texttt{Weight} at
the low end. This turned out to be a car (the
Peugeot 107) with an impossibly low weight of
210 kilograms, so we set that entry to missing.
Rerunning the bixplot on the cleaned data gave the
left panel of Figure~\ref{fig:topgear}.

Among the four variables, \texttt{TopSpeed} and
\texttt{Displacement} are deemed not unimodal and 
fitted by two clusters. Since those variables
have the most structure we then plot them against
each other, with the marginal distributions plotted
in the horizontal and vertical directions by the
bixplot function. We also color coded the points by 
\texttt{DriveWheel}. From this plot we conclude
that there is a group of cars with small engines,
low top speed, and mostly front wheel drive. These
are basically compact city cars. The remaining cars 
are more powerful and faster, with mainly rear 
wheel drive and four wheel drive.

\begin{figure}[H] 
\centering
\includegraphics[width = 1.0\textwidth]
  {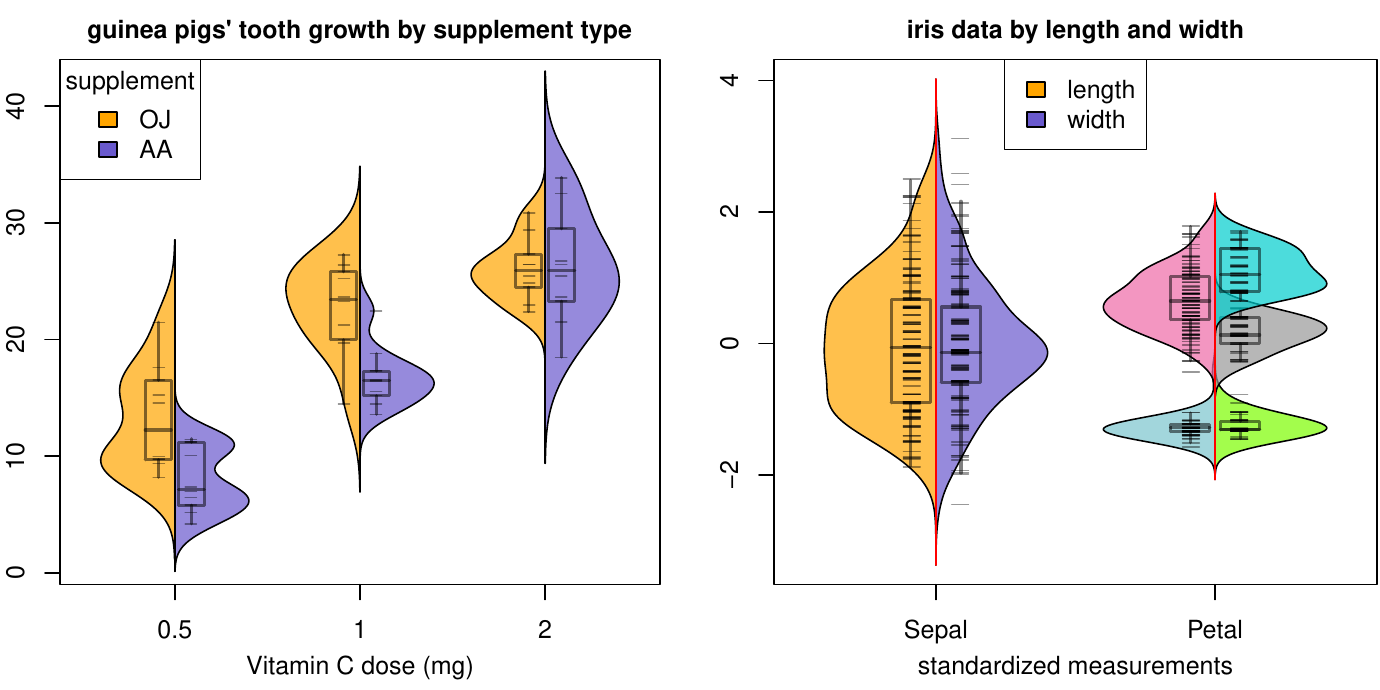}\\
\vspace{2mm}
\caption{Bixplot display of (left) guinea pig's tooth 
growth as a function of vitamin C dose, with two types 
of supplement side by side, orange juice (OJ) and
ascorbic acid (AA); (right) standardized 
variables of the iris data, with length and width side
by side.}
\label{fig:iris+tooth}
\end{figure}

The tooth growth dataset is about an experiment with
60 guinea pigs. Each animal received one of three 
dose levels of vitamin C (0.5, 1, and 2 mg/day) by 
one of two supplements, orange juice (coded as OJ) 
or ascorbic acid (denoted as AA). The response is 
the size of cells responsible for tooth growth. 
The data are available from \textsf{R} as
\texttt{datasets::ToothGrowth}. The left panel of
Figure~\ref{fig:iris+tooth} shows the response in
function of the vitamin C dose, with the two 
supplement types side by side. As there are only 10
animals for each combination, the bixplot function
did not attempt to cluster them. From the display
we see for low and intermediate doses of vitamin C
that orange juice typically had higher responses, 
whereas for the highest dose the median responses
were similar and the variability was lower with
orange juice.

The right hand panel of Figure~\ref{fig:iris+tooth}
shows the same information about the iris data as in
the left panel of Figure~\ref{fig:iris}, but now
displayed with length and width side by side in
four `half' bixplots.

\begin{figure}[H] 
\centering
\includegraphics[height = 8.2cm]
  {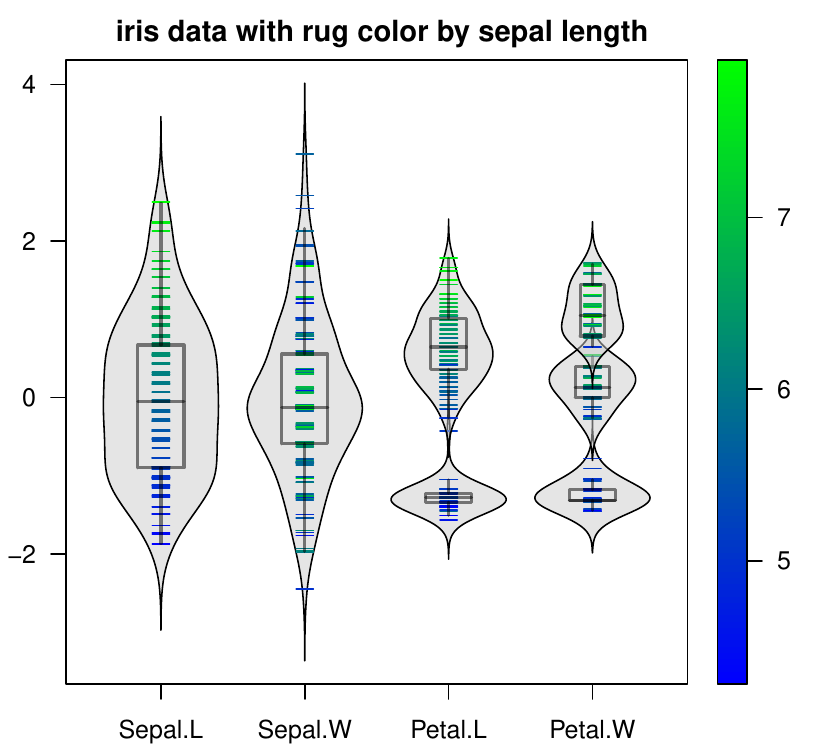}
\includegraphics[height = 8.2cm]
  {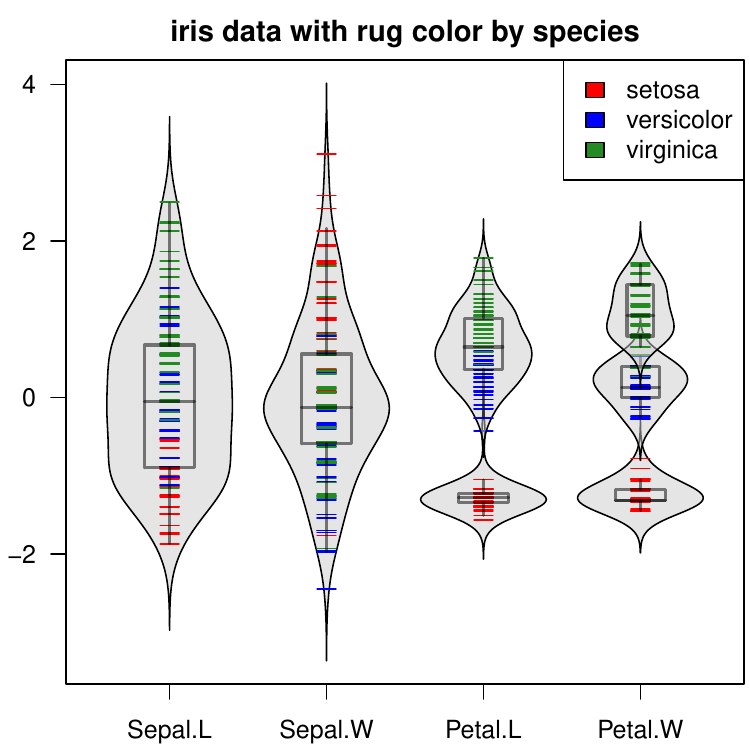}\\
\vspace{3mm}
\caption{(Left) bixplot of standardized variables of 
the iris data, with rug colored by unstandardized
sepal width; (right) rug colored by iris species.}
\label{fig:iris_rug}
\end{figure}

\medskip
The left panel of Figure~\ref{fig:iris_rug}
contains the same information, but moreover the
rug of each variable is colored according to the
first variable, \texttt{Sepal.L}\,. The gradation
is obvious in the bixplot of variable 
\texttt{Sepal.L} itself, but we see that the
rug of \texttt{Petal.L} is colored similarly, 
indicating a positive association between these 
variables. On the other hand there is not much
similarity with the color in \texttt{Sepal.W}\,.

In the right hand panel of Figure~\ref{fig:iris_rug}
the rugs are instead colored according to the 
categorical variable \texttt{Species}. The three
species nicely match the clusters in \texttt{Petal.W}
and \texttt{Petal.L}\,. The colors inside
\texttt{Sepal.L} are roughly similar, whereas
\texttt{Sepal.W} again behaves differently.

\begin{figure}[H] 
\centering
\includegraphics[width = 0.98\textwidth]
  {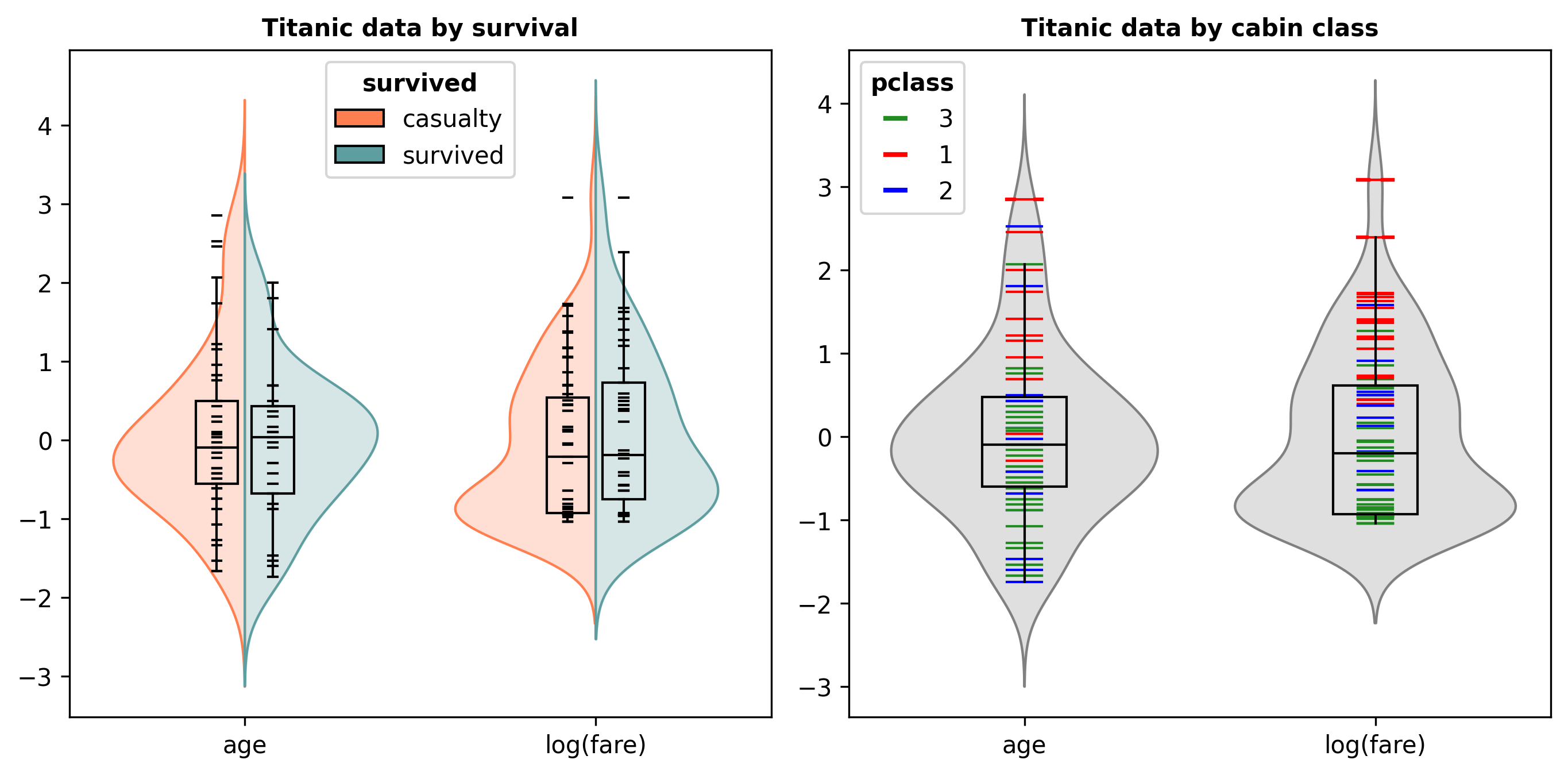}\\
\vspace{2mm}
\caption{(Left) bixplot of standardized 
passenger \texttt{Age} and \texttt{log(Fare)} of 
the Titanic data, with survivors and casualties 
side by side; (right) plot of the same variables 
with the rug colored by cabin class.}
\label{fig:titanic}
\end{figure}

\medskip
The final illustration uses the well-known 
Titanic dataset which contains information 
on the passengers of the RMS Titanic. 
For its history and visualizations see
\cite{Friendly2019}.
The data are freely available from 
\url{https://www.kaggle.com/c/titanic/data}
and the \textsf{R} package \texttt{classmap}
(Raymaekers and Rousseeuw, 2023).
Among the available variables only two are
numeric, the \texttt{Age} of the passenger
and the \texttt{Fare} they paid. Since the
fares are very skewed we decided to visualize
\texttt{log(Fare)} instead. For illustration
purposes we look at the first 100 passengers
in the dataset, and we standardize both
variables. The left panel in
Figure~\ref{fig:titanic} shows their bixplot
display, in which each variable was split
according to the factor \texttt{Survival}.
The results for \texttt{casualty} and for 
\texttt{survived} are shown side by side.
The median age and especially the local 
maximum of its density are higher among the 
survivors than among the casualties. The 
relation between fare and survival is more 
subtle. From the boxes we see that 
the medians of \texttt{log(fare)} 
are similar between the groups, and the 
quartiles are a bit higher for the
survivor group. Here the main difference is
not in the middle of the data.
For the very lowest fares, the density is 
somewhat bigger on the casualty side, and
its local maximum is lower. And for fares 
much above the third quartile, the density 
is a bit bigger on the survivor side. We 
can conclude that there were more casualties 
among passengers with very low fares, and 
a few more survivors among those paying 
very high fares.

We also want to investigate the relation with
another categorical variable, the cabin class.
Since that one has three levels, we cannot do 
so with the side by side mechanism. Instead we
color the rugs by the levels of the cabin class
in the right hand panel of 
Figure~\ref{fig:titanic}. We see that the fare
paid for first class was typically highest,
followed by second class and third class, which
is only natural. But we also see that there is 
a relation between age and cabin class, with 
people of higher age typically occupying
better cabins. This relation seems stronger than 
one might have expected. After seeing this, one
might be tempted to investigate whether such a
pattern has weakened over the last century.

%====================================================
\section{More on the construction of the bixplot}
\label{suppmat:technical}

\medskip
Inside the constrained clustering algorithm, we 
made a small change to the $k$-medoids algorithm.
In its standard version the algorithm restricts
the medoids to cases in the dataset, because
this allows it to deal with datasets that do not 
have coordinates but are given by a matrix of
dissimilarities between cases, or a kernel matrix.
However, here we are in the framework of 
univariate data, so we do know the individual 
data values. For a cluster with an odd number of
members, its medoid is unique and equals its 
sample median. But for clusters with an even 
number of members, its two central data values 
might not coincide, and then the medoid can be 
taken as either of them, so it is not unique. 
However, the median of the cluster is then defined 
as the midpoint of the two central values and is 
therefore unique. Moreover, the objective function 
(the sum of the distances of all cluster members 
to the center) is the same for the median and
for each of the two medoids. Therefore, we 
have opted to use the median in the algorithm
because it is unique. As a side benefit, if the
user would flip the sign of a variable, the
median behaves as expected.

\medskip
In order to be able to interpret clusters as modes we 
need to obtain a partition where each cluster contains
at least \texttt{clusMinN} of unique members,
where \texttt{clusMinN} is given by the user and 
defaults to 3. (Similarly, when a variable has
fewer values than \texttt{clusMinN}, only points are 
drawn.) We have implemented this in a constrained
version of $k$-medoids for given $k$, following ideas of
Bradley et al. (2000) who constructed a version 
of $k$-means 
with this type of constraint. We will provide a sketch of
the algorithm. For simplicity we first assume that the
univariate dataset of size $n$ does not have any ties, so 
each of its values is unique. The algorithm starts with
running an unconstrained $k$-medoid clustering, yielding
a set of $k$ centers denoted as $\mu_g$ for
$g = 1,\ldots, k$. If each cluster already contains
at least \texttt{clusMinN} members, we are done.
If not, we alternate 2 steps. The
first of these is to assign the $n$ cases to the $k$
centers, in such a way that at least \texttt{clusMinN}
cases are assigned to each. For this purpose we will
maintain an $n \times k$ matrix $\bM$ with zero-one
entries $m_{ig}$ for $i=1,\ldots,n$ and groups
$g = 1,\ldots,k$. The entries are membership indicators,
with 
\begin{equation}
m_{i,g} = 
\begin{cases}
1 &\mbox{ if case }$i$\mbox{ is assigned to center }$g$\\
0 &\mbox{ otherwise.}
\end{cases}
\end{equation}
The constraint we need 
to satisfy says that
$\sum_{i=1}^n m_{i,g} \geqslant$ \mbox{\texttt{clusMinN}} 
for each $g = 1,\ldots,k$.
We can write the objective function of $k$-medoids with
another matrix, the $n \times k$ \textit{cost matrix} 
$\bC$ given by its entries
\begin{equation}
  c_{i,g} = |x_i - \mu_g|\,.
\end{equation}
These are the distances of each case to each center,
so the objective function $T$ in the main text equals 
$T = \sum_{i=1}^n \sum_{g=1}^k m_{i,g}\, c_{i,g}$\,.
The first step in the alternation is thus the 
constrained optimization
\begin{equation} \label{eq:lpsolve}
\begin{aligned}
\mbox{minimize}_{\bM}& \quad \sum_{i=1}^n 
  \sum_{g=1}^k m_{i,g}\, c_{i,g}\\
\mbox{subject to}& \quad \sum_{i=1}^n m_{i,g} 
   \geqslant \mbox{\texttt{clusMinN}} 
   \quad \mbox{for}\;\; g = 1,\ldots,k.
\end{aligned}
\end{equation}
This can be seen as a general linear program if we allow 
the memberships $m_{i,g}$ to vary continuously between 0 
and 1, with the constraints $\sum_{g=1}^k m_{i,g} = 1$ 
for $g = 1,\ldots,k$ and $m_{i,g} \geqslant 0$ for
all $i=1,\ldots,n$ and $g=1,\ldots,k$\,. But it is
more efficient to restrict the $m_{i,g}$ to 0 or 1
from the start, making it an integer programming
problem. This enables us to use a technique
developed for transportation problems, implemented as
the function \texttt{lp.transport} in the \textsf{R}
package \texttt{lpSolve} 
(Cs\'ardi and Berkelaar, 2024), and as the 
function \texttt{LpProblem} in the \textsf{Python} 
package \texttt{PuLP} (Mitchell et al, 2011).
Note that the first time we run \eqref{eq:lpsolve}
the objective function $T$ may increase, because the
initial \mbox{$k$-medoid} clustering did not have the
constraint, but afterward it can only decrease the
objective or keep it unchanged.

The other step in the alternation is to recompute the
centers, by
\begin{equation} \label{eq:update}
  \mu_{i,g} \leftarrow \mbox{median}\big\{x_i\;;\; 
  m_{i,g} = 1\big\}\;.
\end{equation}
This is not a linear operation, but it does not have to
be: only the objective in \eqref{eq:lpsolve} needs to be
linear in $\bM$, which it is for any cost matrix $\bC$.
Note that the univariate median minimizes the sum of
distances, so this step will never increase the 
objective function either. It follows that iterating 
the two alternating steps needs to converge after 
sufficiently many iterations, but in the code we have
a limit on the number of iterations anyway, that can
be chosen by the user.

The above description assumed that there were no ties
among the $x_i$ hence they are all unique. If this is
not the case, the optimization \eqref{eq:lpsolve} may
assign tied $x_i$ to different centers as a way to
satisfy the constraint, which is undesirable. Instead we 
first combine tied $x_i$ to cases $y_j$ that are unique, 
and some of which may correspond to more than one $i$. 
We then apply the constrained step \eqref{eq:lpsolve} 
to the $y_j$ to avoid breaking ties. Next we copy the 
resulting memberships of the $y_j$ to the corresponding 
$x_i$ and update the centers in~\eqref{eq:update} using 
all the $x_i$\,, and continue iterating this way.

\spacingset{1.1}

\renewcommand{\refname}{Additional reference} 

\renewcommand{\refname}{References}

\end{document}